\documentclass[conference]{IEEEtran}
\IEEEoverridecommandlockouts
\usepackage{silence}
\usepackage{cite}
\usepackage{amsmath,amssymb,amsfonts}
\usepackage{algorithm,algpseudocode}
\usepackage{graphicx}
\usepackage{textcomp}
\usepackage{xcolor}
\usepackage{tabularx}
\usepackage{stfloats}
\usepackage[compatibility=false]{caption}
\usepackage{subcaption}
\usepackage[left=1.62cm,right=1.62cm,top=1.9cm]{geometry}

\def\BibTeX{{\rm B\kern-.05em{\sc i\kern-.025em b}\kern-.08em
    T\kern-.1667em\lower.7ex\hbox{E}\kern-.125emX}}

\begin{document}

\title{FedBlockHealth: A Synergistic Approach to Privacy and Security in IoT-Enabled Healthcare through Federated Learning and Blockchain\\
}

\author{
\fontsize{9}{11}\selectfont
Nazar Waheed\textsuperscript{1},
Ateeq ur Rehman\textsuperscript{2},
Anushka Nehra\textsuperscript{3},
Mahnoor Farooq\textsuperscript{2},
Nargis Tariq\textsuperscript{2},\\
Mian Ahmad Jan\textsuperscript{4},
Fazlullah Khan\textsuperscript{4},
Abeer Z. Alalmaie\textsuperscript{1},
Priyadarsi Nanda\textsuperscript{1}\\
\fontsize{9}{11}\selectfont
\textsuperscript{1}\textit{School of Electrical and Data Engineering, University of Technology Sydney, Sydney, Australia}\\
\fontsize{9}{11}\selectfont
\textsuperscript{2}\textit{Department of Information Technology, University of Haripur, Haripur, Pakistan}\\
\fontsize{9}{11}\selectfont
\textsuperscript{3}\textit{Department of CS and Engineering, Thapar University Patiala, Patiala, India}\\
\fontsize{9}{11}\selectfont
\textsuperscript{4}\textit{Department of Computer Engineering, Abdul Wali Khan University, Mardan, Pakistan}
}
\maketitle

\begin{abstract}

The rapid adoption of Internet of Things (IoT) devices in healthcare has introduced new challenges in preserving data privacy, security and patient safety. Traditional approaches need to ensure security and privacy while maintaining computational efficiency, particularly for resource-constrained IoT devices. This paper proposes a novel hybrid approach by combining federated learning and blockchain technology to provide a secured and privacy-preserved solution for IoT-enabled healthcare applications. Our approach leverages a public-key cryptosystem that provides semantic security for local model updates, while blockchain technology ensures the integrity of these updates and enforces access control and accountability. The federated learning process enables a secure model aggregation without sharing sensitive patient data. We implement and evaluate our proposed framework using EMNIST datasets, demonstrating its effectiveness in preserving data privacy and security while maintaining computational efficiency. The results suggest that our hybrid approach can significantly enhance the development of secure and privacy-preserved IoT-enabled healthcare applications, offering a promising direction for future research in this field.

\end{abstract}

\begin{IEEEkeywords}
Federated Learning, Blockchain, ElGamal, Privacy Protection
\end{IEEEkeywords}

\section{Introduction}

In healthcare, Internet of Things (IoT) devices like wearables, sensors, and medical equipment have transformed applications and services, enabling remote monitoring, diagnostics, and personalized treatments \cite{razzaque2016middleware}. For effective healthcare, IoT devices must be reliable and accurate, as inaccuracies could result in misdiagnosis or improper treatment. The sensitive patient data collected by IoT devices is vulnerable to various cyber attacks. Thus, robust security measures are essential to protect patient privacy and ensure safety. Regulatory requirements, such as HIPAA and GDPR, further emphasize the need for robust solutions \cite{roman2018fog}.

Traditional centralized machine learning approaches have various drawbacks, including higher communication costs, battery consumption, and potential security risks \cite{konevcny2016federated}. Federated Learning (FL), a distributed machine learning approach, has emerged as an alternative that enhances user privacy by training models over remote devices or data centers without sharing the raw data \cite{mcmahan2017communication}. However, FL is vulnerable to poisoning attacks, and attackers can recover data from gradients \cite{Nazar, hitaj2017deep}. Integrating FL with blockchain has been explored in healthcare \cite{azaria2016medrec, jiang2018blockchain}, but these studies lack comprehensive solutions addressing privacy and security while maintaining computational efficiency.

FL faces security vulnerabilities, model consistency and accuracy, limited network bandwidth, and data imbalances between clients \cite{yang2019federated}. We propose FedBlockHealth, a novel hybrid approach combining FL and blockchain (BC) technology for secure and privacy-preserved IoT-enabled healthcare applications to address these vulnerabilities. Our approach leverages a public-key cryptosystem for local model update semantic security and BC for decentralized data storage, management, and access control. FL ensures secure model aggregation without sharing sensitive patient data, maintaining privacy, security, and computational efficiency.

Research has addressed FL challenges, such as non-IID data distributions \cite{smith2017federated}, global model convergence \cite{sattler2019robust}, and the client device and data heterogeneity \cite{li2020federated}. Differential privacy has been introduced \cite{abadi2016deep}, and applied to FL \cite{triastcyn2019federated}. Blockchain has been used to secure patient data \cite{ateeqBC1} but with centralized machine learning for disease detection.

Our FedBlockHealth framework offers vital contributions to privacy-preserving IoT-enabled healthcare. These contributions include

\begin{itemize}
    \item \textbf{Hybrid Approach:} An Algorithm is proposed to  combine FL and BC technology to  address privacy and security challenges in IoT-enabled healthcare scenarios. Smart contracts are designed for all clients to data on BC.  This hybrid approach ensures the integrity and privacy of patient data while maintaining computational efficiency. 

    \item \textbf{Semantic Security:} To secure the communication between clients and server, the Elgamal public-key cryptosystem is employed in our framework to provide semantic security for local model updates and ensures that sensitive patient information remains private during the FL process. 


    \item \textbf{Performance Evaluation:} Our proposed  CNN framework for FL is evaluated using the EMNIST dataset.  The approach maintains privacy and security while ensuring computational efficiency.  Results demonstrate its effectiveness compared to traditional ANN and CNN models by achieving 99.9$\%$ accuracy  and a loss of 0.01$\%$. This is the result of reducing the number of  layers and adding a batch Normalization layer, which  alternatively  decreased complexity and the use of Feedforward and Backpropagation approach. 

    \item \textbf{Scalability and Applicability:} Our framework FedBlockHleath is designed to be scalable and applicable to a wide range of healthcare scenarios, paving the way for future research and development in secure and privacy-preserving IoT-enabled healthcare applications.
\end{itemize}
In this paper, we discuss the problem statement and system architecture in Section 2, followed by the methodology in Section 3. Experimental results and analysis are presented in Section 4, and finally, we conclude the study and suggest future research in Section 5.
%
\section{Proposed System Architecture}



\begin{figure}[hbt!]
\centering
\includegraphics[width=0.475\textwidth]{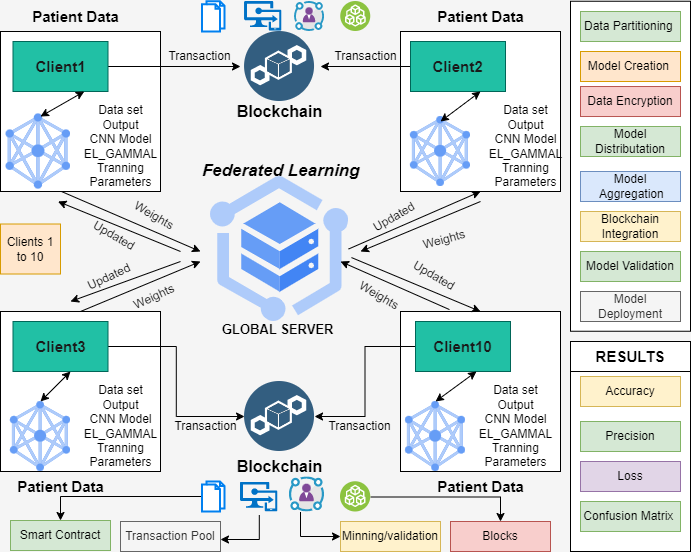}
\captionsetup{justification=centering}
\caption{ The proposed FedBlockHealth model, where clients data are stored in blockchain and local model weights are shared with global server.  }
\label{fig:SystemArchitecture}
\end{figure}

To circumvent the problem of data privacy and security, in this paper, we proposed the FedBlockHealth model illustrated in Fig. \ref{fig:SystemArchitecture}, in which patient data is received through IoT devices and stored in the hospital database represented by the client. Each client has a locally trained CNN model shared by the global server. Each client trains independently, and the updated weights, encrypted through the EL-Gammal approach, are shared with the global server for aggregation. On the other hand, each client sends model updates to the BC for data storage. Note that a smart contract is designed for each client before the transaction in model updates are forwarded to the transaction pool for validation. After completing the validation process, the system stores the transaction in blocks, and authorized users from the BC network can only access the data.

\subsection{System Architecture}

The proposed system architecture, FedBlockHealth, comprises a Key Generation Centre (KGC), a server, and multiple participants.

\subsubsection{KGC}

The KGC is a trusted entity responsible for system setup, public parameter generation, and private key distribution to each participant and the server. The KGC is fully trusted and does not collide with any other entities.

\subsubsection{Server}

The server is a shared location for securely aggregating encrypted local gradients from all participating clients and distributing updated global parameters. In encrypted computing techniques, an adversarial setting is often assumed, where all entities, except the KGC, follow the protocol precisely while attempting to infer sensitive data from the training data.

\subsubsection{Clients}

Each client represents a hospital that possesses a health-related dataset and a copy of the trained model shared by the global server. During learning, clients train their local models on their private datasets and share local gradients/weights with the server. Malicious clients may attempt to communicate with the server for nefarious purposes, necessitating measures to prevent such activities and protect the data privacy of legitimate clients. Hence, the EL-Gammal encryption approach is used to secure communication between clients and the global server. On the other hand, smart contracts are created for each client to communicate with the BC network. Figure \ref{fig:SystemArchitecture} illustrates the proposed system architecture.

\subsubsection{Federated Learning}
Federated learning is a distributed machine learning technique allowing multiple parties to train a model collaboratively without sharing their raw data. In our system model, we assume that multiple healthcare providers (clients) are participating in the FL process to improve the accuracy of the Convolutional Neural Network (CNN) model.
Each healthcare provider maintains their datasets of medical images and trains a local CNN model using FL. The local model weights are encrypted using an ElGamal cryptosystem before being sent to the central server. The central server aggregates the encrypted model weights and updates the global CNN model. The updated global model weights are then sent back to the clients for further training. The process continues iteratively until the model converges to a satisfactory accuracy level as reflected in Algorithm \ref{CNNalgorithm}. FL allows stakeholders to improve their models by leveraging the network's collective intelligence without sharing their personal data.

\subsubsection{Blockchain Network}

The proposed system uses BC technology to manage access control and provide an immutable record of the training process in FL in healthcare. The system initially designs smart contracts for each client and then transfers the model updates to a smart contract for conversion to hash as shown in Algorithm \ref{smart_contract}. Afterward, the system forwards the hash to the transaction pool, and the miners pick it for mining. Once the miners have picked the hash, they create blocks. The authorized clients can retrieve the data stored in blocks through their smart contract, as depicted in Algorithm \ref{algorithm_BC}. This approach enables the system to manage access to the distributed CNN model transparently and securely while ensuring the integrity of the training process.



\subsection{Federated Learning with Blockchain Integration}

We can transform healthcare data management, analysis, and use by integrating BC technology and FL. Using BC technology, we can manage patient data securely and transparently while employing FL to develop predictive models for disease diagnosis or personalized treatment plans. We can achieve the integration of BC technology and FL in healthcare through the following steps:
\begin{enumerate} 
\item Data partitioning: The healthcare data is partitioned among multiple hospitals or healthcare institutions to ensure data privacy and security. Each hospital holds its data locally.

\item Model creation: A central server creates a CNN model, which trains the data from all the participating hospitals through the federated averaging algorithm presented in Algorithm \ref{CNNalgorithm}.

\item Data encryption: The hospital data is encrypted using a cryptographic algorithm ElGamal discussed in the following subsection, adding an extra layer of security and privacy.

\item Model distribution: The central server receives the encrypted data and distributes the CNN model to all participating hospitals.

\item Local model training: Each hospital trains the CNN model on their local encrypted data using FL, which involves multiple rounds of model training and aggregation of model updates.

\item Model aggregation: The hospitals send their encrypted model updates to the central server, which aggregates the updates to create a global model as presented in Algorithm \ref{CNNalgorithm}.

\item Blockchain integration: The updated models of clients are stored on a BC, ensuring the model updates' transparency and immutability.

\item Model validation: A third-party auditor validates the global model to confirm that it meets the necessary accuracy and security standards.

\item Model deployment: The validated global model is then deployed back to the participating hospitals for local inference on new data.
\end{enumerate}

\begin{algorithm}[tbh!]
\caption{Smart Contract }
\label{smart_contract}
\begin{algorithmic}[1]
\State  {\bf INPUT Client Registration }
\If{Registration == successful}
   \State Check data from the Global server 
\Else
  \State Register client as a New Client 
   \State U $\leftarrow$ Client
        \State Store Data on Blockchain 
        \State Encrypt data Using  Sha256 Algo
        \State {\bf OUTPUT} generate Application Binary Interface of Contract
        \State generate Byte code of Contract 
        \State Decrypt data Using  Sha256 Algorithm 
\EndIf
\end{algorithmic}
\end{algorithm}

\begin{algorithm}[tbh!]
\label{BC_OP}
\caption{Operation of Blockchain  }
\begin{algorithmic}[1]
\State  {\bf INPUT Client Data }
\State Client Data 
\State Data added to Smart Contract 
\State Data Converted using SHA256 Hashing Algorithm
\State Data transfer to transaction Pool
\State Minners minne the transaction 
\State Transaction are converted into blocks
\State {\bf OUPUT} Client data 
\Comment{Only Authorized person access the data}
\end{algorithmic}
\end{algorithm}

\begin{algorithm}
\caption{Blockchain integration with Federated Learning}
\label{algorithm_BC}
\textbf{Initialize:} Initialize blockchain network
\\  Create a smart contract for FL Server and Clients
\begin{algorithmic}[1]
\State Client-Side Operations:
\For{for each client U in the pool}
\State Connect to the blockchain network
\State Register as a client with the smart contract
\State Load pre-trained CNN model weights
\State The FL Agorithm will perform its operation
\EndFor
\State \textbf{Server-Side Operation}
\State Connect to the blockchain network
\State Register as the server with the smart contract
\State Collect local gradients from clients
\State The FL Agorithm will perform its operation
\State \textbf{Blockchain Operations}
\State Record transactions on the blockchain
\State Maintain a tamper-proof record of all transactions
\State Ensure transparency and accountability 


\end{algorithmic}
\end{algorithm}

\subsection{Feedforward and Backpropagation}

Clients discretize their model updates, add discrete Gaussian noise, and submit them for modular secure summation. This comprehensive end-to-end system employs the ElGamal encryption scheme. To perform a forward pass through the network, we use the following iterative formula to compute each neuron in the subsequent layer \cite{steinke2021distributed}:
	\vspace{-0.1cm}
\begin{equation}
a^l = \sigma \times (W \times a^{(l-1)}+b)
\end{equation}
	\vspace{-0.1cm}
Backpropagation efficiently computes gradients, and the optimizer uses these gradients to train the neural network:
	\vspace{-0.1cm}
\begin{equation}
w^{t+1} = w^t - n \times \nabla_{w} \times L(D^t,w^t)
\end{equation}

The equation represents the Feedforward function $f(x, w) = y$, where $w$ is the parameter vector, and the training dataset is $D = \{(x_i, y_i); i \in L\}$.
$L$ represents the loss function, and backpropagation is defined as:

\begin{equation}
\frac{1}{|D|} \sum_{(x_i, y_i) \in D} \ell (y_i, f(x_i,w))
\end{equation}

Training continues until the loss function reaches an optimal minimum value. Following this approach, the proposed system architecture effectively addresses distributed learning challenges, ensuring data privacy and successfully training a high-quality centralized model.

\section{METHODOLOGY}

In order to enhance data privacy, this study investigates the performance of FL on the Extended Modified National Institute of Standards and Technology (EMNIST) datasets by utilizing a Convolutional Neural Network (CNN). FL on EMNIST datasets using a CNN can address the challenge by training the model on local data held on different healthcare devices while keeping the data decentralized and preserving the privacy of individual patients. The architecture of the employed CNN comprises four convolutional layers, each succeeded by a max-pooling layer and an additional two fully connected layers within the hidden layer structure. Given the flexibility of the convolutional layer, the subsequent section presents the proposed CNN model tailored explicitly for FL applications. Only retrain the pre-trained model and the fully connected layers after using previously trained convolutional layers. We examined the performance of FL on the EMNIST datasets using a CNN.

\begin{algorithm}
\caption{Federated Learning for CNN}
\label{CNNalgorithm}

\textbf{Initialize:} Training dataset, Test dataset, libraries  \\
\textbf{Input:} Training dataset, Test dataset, EMNIST images \\
\textbf{Output:} Trained and secured data
\begin{algorithmic}[1]

\State Do Initialization procedures
\State initialize blockchain
\State import np $\rightarrow$ linear algebra
\State import pd $\rightarrow$ data processing CSV file
\State Upload $D_T$, $D'_T$ as csv file
\State From $sklearn.model\_selection$ import $D_T$, $D'_T$
\State Create CNN Model
\For{for each client u in the network:}
    \State generate key pair (public key, private key) for client u
    \State Store public key on the Blockchain
    \EndFor
\State Import $tff$ $\rightarrow$ for federated learning model
\For{Each iteration $n = 1:N $}
    \For{Each image-training step $l = 1:L$}
        \For{Each iteration $ k =1:K $}
            \State Select and go to next pixel
            \State Update pixels Sigmoid
        \EndFor
    \State Update visited pixels Relu
    \EndFor

    \For{Every client $ u $}
        \State complex equation
        \State Collect client contribution
    \EndFor

    \State complex equation
    \State Distributed among clients
\EndFor
\State Accuracy check:
\For{each iteration epoch}
    \Statex \hspace{1.5em} $n=70$, Num\_batch$=10$,
    \Statex \hspace{1.5em} Num\_clients$=10$,
    \Statex \hspace{1.5em} Shuffle\_Buffer$=100$,
    \Statex \hspace{1.5em} Prefetch\_Buffer$=10$
\EndFor
\end{algorithmic}
\end{algorithm}

In order to improve the accuracy of our model, we have introduced two additional hidden layers, which enable the extraction of more complex features from the input data. We have utilized the stochastic gradient descent (SGD) optimizer to optimize the model's performance further. The SGD optimizer iteratively refines the model's parameters to minimize the discrepancy between the predicted and actual outputs. This is accomplished by computing the gradient of the loss function with respect to the parameters and updating them in the opposite direction of the gradient, ultimately determining the optimal parameters for the model. As a result, the model can more accurately predict the output.
Furthermore, securing the communication between clients and the global server is vitally important; therefore, we use the El-Gamal Multiplicative Cryptosystem approach. El-Gamal Multiplicative Cryptosystem is preferred over other techniques because it provides both confidentiality and data integrity during transmission. Unlike other techniques, it simultaneously encrypts plaintext and signature generation, ensuring that the data remains secure and unaltered during transmission. Additionally, ElGamal encryption relies on mathematical problems that are computationally difficult to solve, making it a more secure approach to encryption \cite{elgamal1985public}.

\subsection{El-Gamal Multiplicative Cryptosystem}

ElGamal encryption is a public-key cryptography algorithm that is based on the Diffie-Hellman key exchange \cite{steinke2021distributed}. In the context of federated learning, it can be used to encrypt the model parameters before they are sent from the client devices to the central server for aggregation. This helps to ensure that the model parameters remain secure and confidential during the transmission.

The ElGamal encryption algorithm comprises three parts: \begin{itemize}
\item \textbf{Key generation}: In this step, a user generates a public-private key pair. The public key encrypts, while the private key decrypts. \item {\bf{Encryption:}} To encrypt the model parameters, the client device selects a random value known as the session key. The client uses the session key to encrypt the model parameters with ElGamal encryption. Then, the encrypted session key and model parameters are sent to the central server for aggregation. \item {\bf{Decryption}}: The central server uses its private key to decrypt the encrypted session key and the encrypted model parameters. The decrypted session key decrypts the encrypted model parameters. \end{itemize} The security of the ElGamal encryption algorithm relies on the difficulty of the Discrete Logarithm Problem (DLP), which involves finding the exponent $x$ in the equation $g^x \bmod p = h$. This problem is computationally complex, making ElGamal encryption secure against attacks from hackers.
In summary, using the ElGamal cryptosystem in federated learning with the CNN model adds a security layer to the model parameters during transmission. This ensures the privacy and confidentiality of the model parameters, even when transmitted over a public network. The El-Gamal Multiplicative Cryptosystem Algorithm is presented in \ref{alg:algorithm2}, while its mathematical proof will appear in the extended version of this model.

\begin{algorithm}
\caption{Pseudo-code of the ElGamal technique}
\label{alg:algorithm2}
\begin{algorithmic}[1]
\State \textbf{Input:} Exponential ElGamal
\State \textbf{Output:} Messages are encoded by exponentiation
\State Give Value $v_1 = 4, v_2 = 5$
\State $v_1 \gets$ generator.selfApply($4$)
\State $v_2 \gets$ generator.selfApply($5$)
\State $c_1 \gets$ ElGamal.encrypt(Public Key, $v_1$)
\State $c_2 \gets$ ElGamal.encrypt(Public Key, $v_2$)
\State Combine $\gets c_1.\text{apply}(c_2)$
\State Results $\gets$ ElGamal.decrypt(Private Key, Combine)
\State Calculate $v = v_1 \cdot v_2$ in message space:
\Statex \hspace{2.5em} generator.selfApply($v = v_1 \cdot v_2$)
\State Print results
\State End
\end{algorithmic}
\end{algorithm}

\section{Performance Evaluation and Simulation}
This section aims to evaluate the performance of the proposed FL system using the EMNIST datasets with a Convolutional Neural Network (CNN). We compare the privacy-enhanced system to the non-private baseline and discuss the trade-offs between privacy, model complexity, and accuracy.

\subsection{Dataset and Model Architecture}

We use the EMNIST dataset, which consists of 28x28 gray-scale images of handwritten digits. We divide the dataset into 71,039 training samples, 14,799 testing samples, and 17,760 validation samples. We further partition the data among ten clients for distributed training.

\vspace{0.31in}
\begin{table*}[t]
\centering
\caption{Model details and performance comparison}
\begin{tabular}{|l|l|l|l|}
\hline
 & Base paper CNN model details & CNN Model we apply on our dataset & ANN Model we apply on our dataset \\ \hline
Training data set & 60,000 & 71,039 & 71,039 \\ \hline
Testing dataset & 10,000 & 14,799 & 14,799 \\ \hline
Validation dataset & - & 17,760 & 17,760 \\ \hline
Model & CNN & CNN & ANN \\ \hline
Accuracy Level & 99.03\% at 150 Epoch & 99.99\% at 40 Epoch & 68.91\% at 70 Epoch \\ \hline
\end{tabular}
\label{tab:model_comparison}
\end{table*}

The CNN model used for evaluation consists of four convolutional layers, max-pooling layers, two fully connected layers and a batch normalization layer. The convolutional layers extract essential features from the input images, while the deeper layers identify more complex patterns. Similarly, we use flattened and dense layers to convert the output of the convolutional layer into a single-dimensional vector and perform classification, respectively.

\subsection{Training and Evaluation}

\begin{figure}[hbt!]
\vspace{-0.2cm}
\centering
\includegraphics[width=0.80\columnwidth]{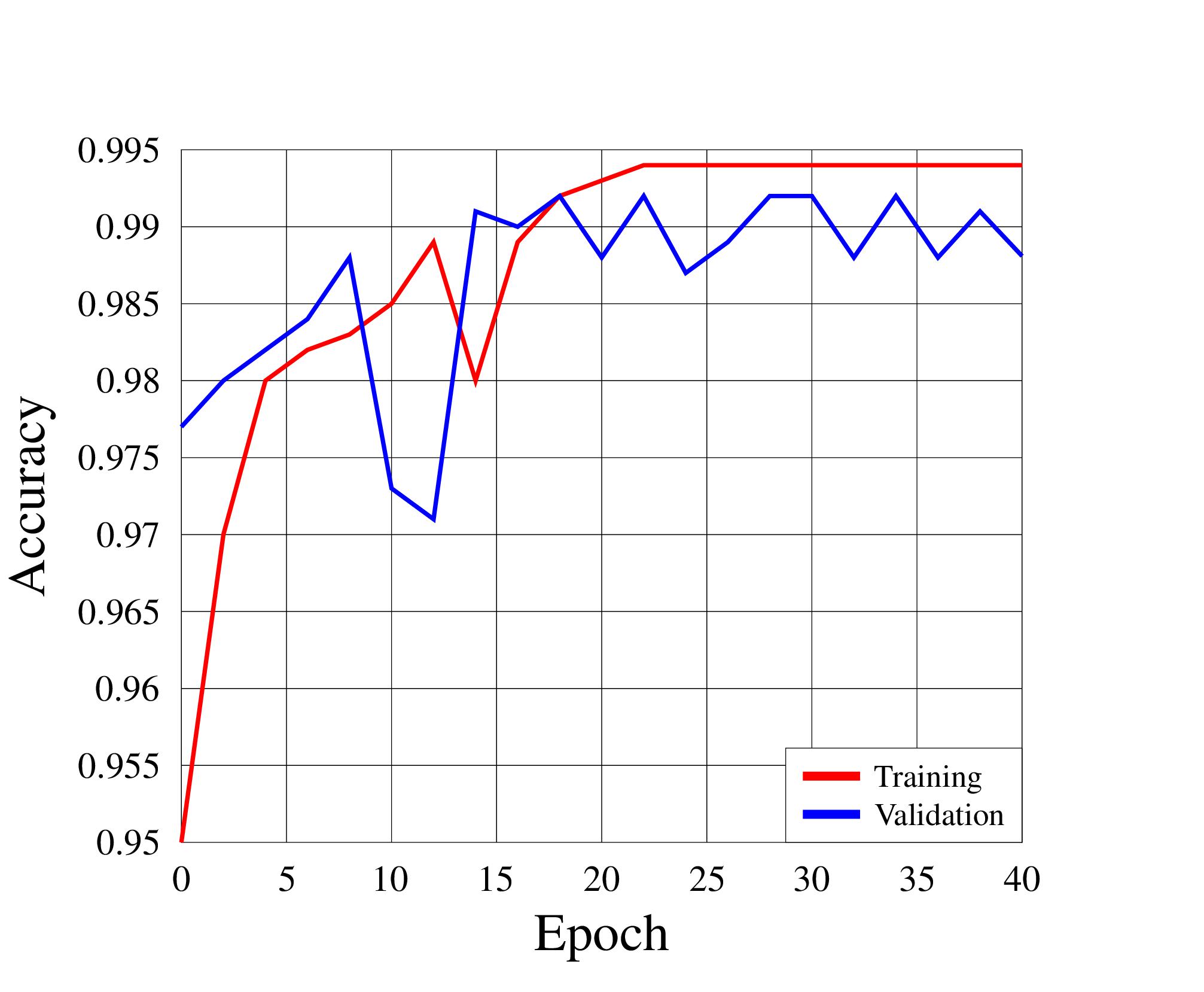}
\caption{ Accuracy of FedBlockHealth  based CNN Model.}
\vspace{-0.2cm}
\label{fig:accu}
\end{figure}

\begin{figure}[hbt!]
\centering
\includegraphics[width=0.80\columnwidth]{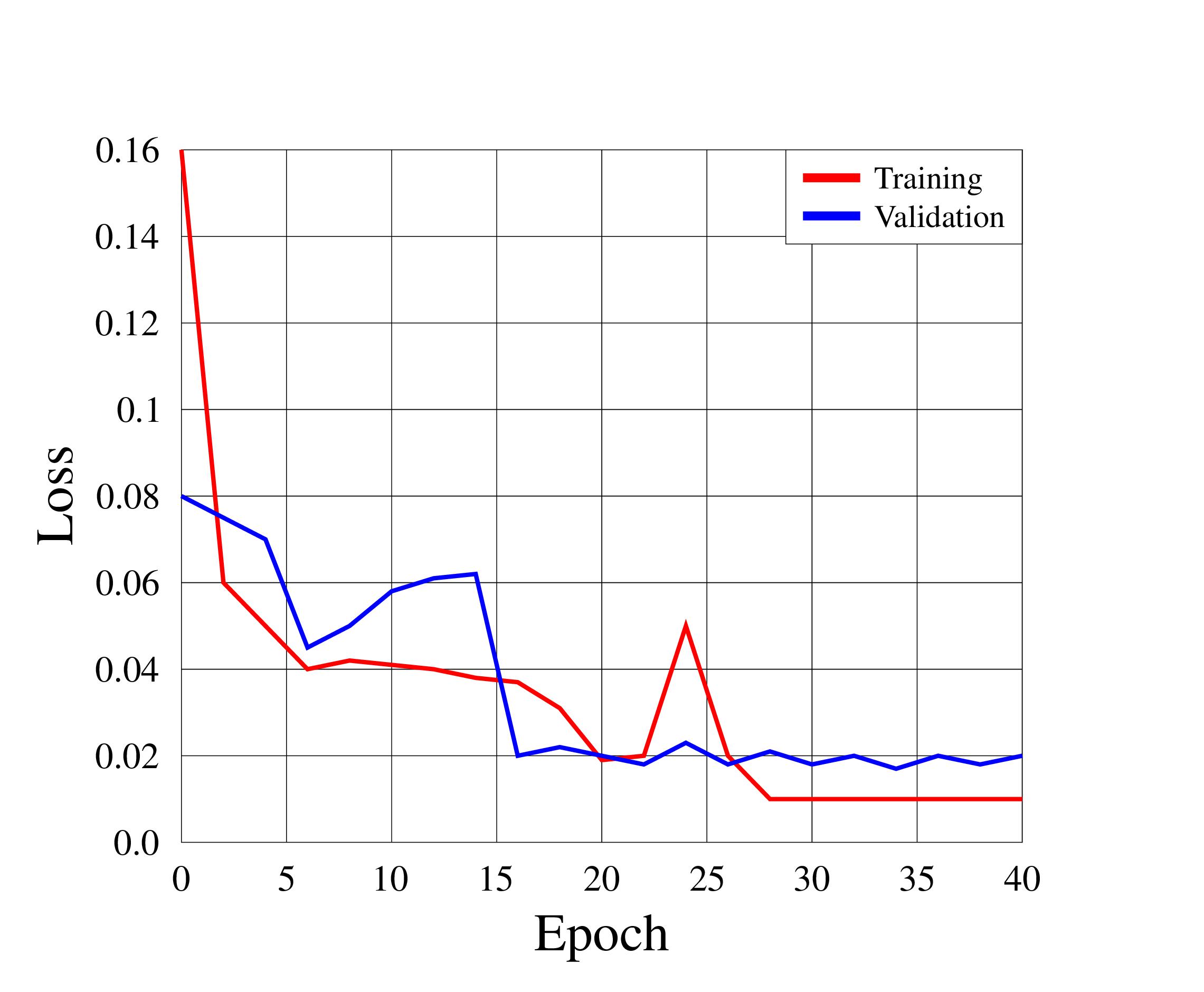}
\caption{Loss of FedBlockHealth based CNN Model.  }
\vspace{-0.2cm}
\label{fig:los}
\end{figure}

During the training phase, the clients train their local models on their respective datasets and share the local gradients with the central server. The server aggregates the gradients, updates the global model, and broadcasts it back to the clients. This process repeats until the loss function converges to an optimal value.
Figures \ref{fig:accu} and \ref{fig:los} show the accuracy and loss graphs of the CNN model applied to the EMNIST dataset. After 70 epochs, we achieved an accuracy of 95.57\% and a loss value of 1.4.

\subsection{Comparison with Baseline and Alternative Models}

We compare the performance of our proposed system with the non-private baseline model and an alternative   Artificial Neural Network (ANN) model. The baseline CNN model achieved an accuracy of 99.03\% after 150 epochs, while our privacy-enhanced CNN model reached an accuracy of 99.99\% after 40 epochs. The ANN model, on the other hand, achieved an accuracy of 68.91\% after 70 epochs. This is due to the reduced complexity of the model by limiting the hidden layers, using the batch norm layer for faster training, and using the Feedforward and Backpropagation approach. The performance is reflected in Table \ref{tab:model_comparison} summarizes the model details and performance comparisons.

The results indicate that the proposed privacy-enhanced FL system with a CNN model achieves competitive performance compared to the non-private baseline. Although the accuracy is slightly lower, the model ensures privacy preservation and employs a less complex encryption technique. In contrast, the ANN model performs significantly worse, demonstrating the importance of using appropriate model architectures for the specific problem domain.

\section{Conclusion}

This study presented a privacy-enhanced federated learning (FL) system, incorporating blockchain and smart contracts, using a convolutional neural network (CNN) for distributed training on the EMNIST dataset. The system effectively balances data privacy preservation and model performance, making it a suitable solution for sensitive data tasks in IoT-enabled healthcare applications.
Our evaluation demonstrates that the privacy-enhanced CNN model achieves 99.99\% accuracy.  
We employed ElGamal encryption to maintain anonymity while enabling computation in the ciphertext space. This method ensures privacy preservation and utilizes a less complex encryption technique. Additionally, integrating blockchain technology and smart contracts enhance the integrity and security of the system.
Future work will involve using high-dimensional datasets and exploring more complex neural network models to enhance accuracy and efficiency in privacy-preserving IoT-enabled healthcare applications.

\bibliographystyle{IEEEtrans}
\bibliography{references}

\end{document}